\documentstyle[12pt,epsf]{article} 
\newcommand{\be}{\begin{equation}} 
\newcommand{\en}{\end{equation}}
\newcommand{\bea}{\begin{eqnarray}}
\newcommand{\ena}{\end{eqnarray}}

\newcommand{\hbo}{\hbox to 1 true cm {\hfill } }

\def\dslash{\partial\kern-.6em\slash}
\def\kslash{k\kern-.5em\slash}
\def\pslash{p\kern-.4em\slash}
\def\Dslash{D\kern-.6em\slash}
\def\Vslash{V\kern-.7em\slash}
\def\vslash{v\kern-.5em\slash}
\def\rslash{r\kern-.5em\slash}
\def\qslash{q\kern-.5em\slash}

\begin{document} 
\vglue 1truecm
  
\vbox{ UNITU-THEP-19/97 
\hfill October 20, 1997
}
  
\vfil
\centerline{\large\bf Confinement and scaling of the vortex vacuum } 
\centerline{\large\bf of SU(2) lattice gauge theory$^*$ } 
  
\bigskip
\centerline{ Kurt Langfeld, Hugo Reinhardt and Oliver Tennert }
\vspace{1 true cm} 
\centerline{ Institut f\"ur Theoretische Physik, Universit\"at 
   T\"ubingen }
\centerline{D--72076 T\"ubingen, Germany}
  
\vfil
\begin{abstract}

The magnetic vortices which arise in SU(2) lattice gauge theory in 
center projection are visualized for a given time slice. 
We establish that the number of vortices piercing a given 2-dimensional 
sheet is a renormalization group invariant and therefore physical 
quantity. We find that roughly $2$ vortices pierce an area of 
$1\, \hbox{fm}^2 $.

\end{abstract}

\vfil
\hrule width 5truecm
\vskip .2truecm
\begin{quote} 
$^*$ Supported in part by DFG under contract Re 856/1--3. 
\end{quote}
\eject

\centerline{ \bf 1. Introduction \hfill }
\medskip 

The conjecture~\cite{tho76} that confinement is realized as a dual 
Meissner effect by a condensate of magnetic monopoles recently received 
strong support by lattice calculations which show that 
in certain gauges the magnetic monopole configurations account 
for about 90\% of the string tension~\cite{r2,pol95,r5}. 
The dominance of magnetic 
monopoles is most pronounced in the so-called 
Maximum Abelian gauge, where the influence of the charged components 
of the gauge field is minimized and which is a precursor of the 
Abelian projection, where the charged components of the gauge fields 
are neglected. The existence of magnetic monopoles is not restricted 
to the Maximum Abelian gauge. Monopoles occur, if the gauge fixing 
procedure leaves an U(1) degree of freedom 
unconstrained~\cite{kro87,la97}. The monopoles carry magnetic charge 
with respect to this residual U(1) gauge freedom. 

\vskip 0.3cm 
If confinement is realized as a dual Meissner effect, i.e. by a 
condensation of the magnetic monopoles, the residual U(1) degree of 
freedom should be broken in the confining phase by the Higgs mechanism. 
This suggests that the relevant infrared degrees of freedom may be 
more easily identified in a gauge where the residual gauge is fixed. 
The U(1) gauge symmetry is explicitly broken 
in the Maximum Center gauge of the lattice theory, which has been 
implemented on top of the Maximum Abelian gauge~\cite{deb96}. 
The Maximum Center gauge preserves a residual $Z_2$ gauge symmetry. 
Analogously to the Abelian projection, one has studied the center 
projection of SU(2) lattice gauge theory, i.e. the projection of 
SU(2) link variables (in Maximum Center gauge) onto $Z_2$ center 
elements. The important finding in~\cite{deb96} (by numerical studies) 
has been a significant center dominance, i.e. the center projected 
links carry most of the information about the string tension of the 
full theory~\cite{deb96}. This naturally leads 
to the conjecture that the field configurations which are eliminated 
by the center projection are not relevant for confinement. 
Furthermore, the numerical calculations also reveal that 
the signal of the scaling behavior of the string tension is much clearer 
in center projection than in Abelian projection. 

\vskip 0.3cm 
The center projection gives rise to vortices which are defined by a 
string of pla\-quettes. Plaquettes are part of the string, if 
the product of the corresponding center--projected links is 
$(-1)$ (for details see~\cite{deb96}). The results in~\cite{deb96} 
indicate that these $Z_2$-vortices are the ''confiners'', i.e. the 
field configurations relevant for the infra-red behavior of the theory. 
This result supports a previous picture by 't~Hooft~\cite{tho79} and 
Mack~\cite{mack} in which the random fluctuations in the numbers 
of such vortices linked to a Wilson loop explain the area law. 

\vskip 0.3cm 
In this letter, we will further investigate the properties of the vortices 
introduced by center projection on top of Abelian projection as 
considered by Del Debbio et al.~in~\cite{deb96}. We will 
present a visualization of these vortices in coordinate space at a given 
time slice. The splitting of the vortices into branches is briefly 
addressed. By calculating the vortex distribution at several values 
of the inverse coupling $\beta $, we will establish that the string 
density is renormalization group invariant and therefore a physical 
quantity. On the average, we will find two vortices piercing an area 
of $1 \, \hbox{fm}^2$.

\vskip 0.3cm 
\centerline{ \bf 2. The vortex vacuum in SU(2) lattice gauge theory \hfill } 
\medskip

\begin{figure}[t]
\parbox{6cm}{ 
\hspace{1cm} 
\centerline{ 
\epsfxsize=6cm
\epsffile{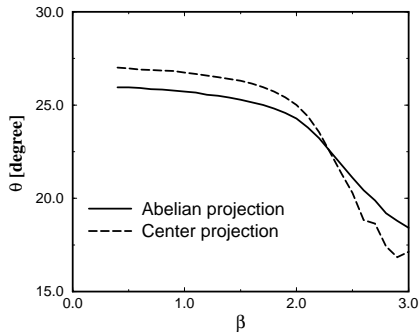} 
}
} \hspace{1cm}
\parbox{7cm}{ 
\caption{ The angle $\theta $ is a measure for the error induced 
   by Abelian and center projection, respectively. } 
} 
\label{fig:1} 
\end{figure} 
Center projection was defined 
in~\cite{deb96} on top of Abelian projection. The Maximum Abelian 
gauge~\cite{tho76} makes a link variable $U_\mu (x)$ as 
diagonal as possible, and Abelian projection replaces a link variable 
\be 
U_\mu (x) \; = \; \alpha _0(x)  \; + \; i \vec{\alpha } (x) 
\; \vec{\tau } \; , \hbo 
\alpha _0^2 + \vec{\alpha }^2 =1 
\label{eq:1}
\en 
by the Abelian link variable 
\be 
A \; = \; \frac{ \alpha _0(x)  \; + \; i \alpha _3 (x) 
\; \tau ^3 }{ \sqrt{ \alpha _0^2 + \alpha _3^2 } } \; = \; 
\cos \theta (x) \; + \; i \sin \theta (x) \; \tau ^3 \; . 
\label{eq:2}
\en 
In order to quantify the error done by this projection, we 
introduce an angle $\theta ^{A} $ by 
\be 
\hbox{tan} \, \theta ^A \; = \; \frac{ 
\left\langle \sqrt{ \alpha _1^2 + \alpha _2^2 } \right\rangle }{ 
\left\langle \sqrt{ \alpha _0^2 + \alpha _3^2 } \right\rangle } \; , 
\label{eq:3} 
\en 
which measures the strength of the charged components relative 
to the strength of the neutral ones. The brackets in (\ref{eq:3}) 
indicate that the lattice average of the desired quantity is taken. 

Center projection is then defined by assigning to each (Abelian) link 
variable a value $\pm 1$ according the rule $A (x) \rightarrow \hbox{sign }
(\cos \theta (x) ) $. The error which is induced 
by the center projection can be measured by $\theta ^C$, i.e. 
\be 
\hbox{cos} \, \theta ^C \; = \; \sqrt{ \left\langle \cos^2 \theta 
\right\rangle } \; . 
\label{eq:4} 
\en 

Our numerical result for the angles $\theta ^A$ and $\theta ^C$ 
is shown in figure 1 as a function of the inverse coupling 
$\beta $. In the scaling window, i.e. $\beta \in [2,3]$, the 
$\theta $-angle is of the order $20^0$ degrees in either case. 
The relative error of a measured quantity on the lattice 
induced first by Abelian and  subsequently by center projection 
is generically given by $\sin \theta ^{A} + \sin \theta ^{C} \approx 
75 \%$. Hence, the charged components of the link variable, i.e. $a_1$ and 
$a_2$, are not small compared with the neutral components, i.e. 
$a_0$ and $a_3$. 

\begin{figure}[t]
\centerline{ 
\epsfxsize=7cm
\epsffile{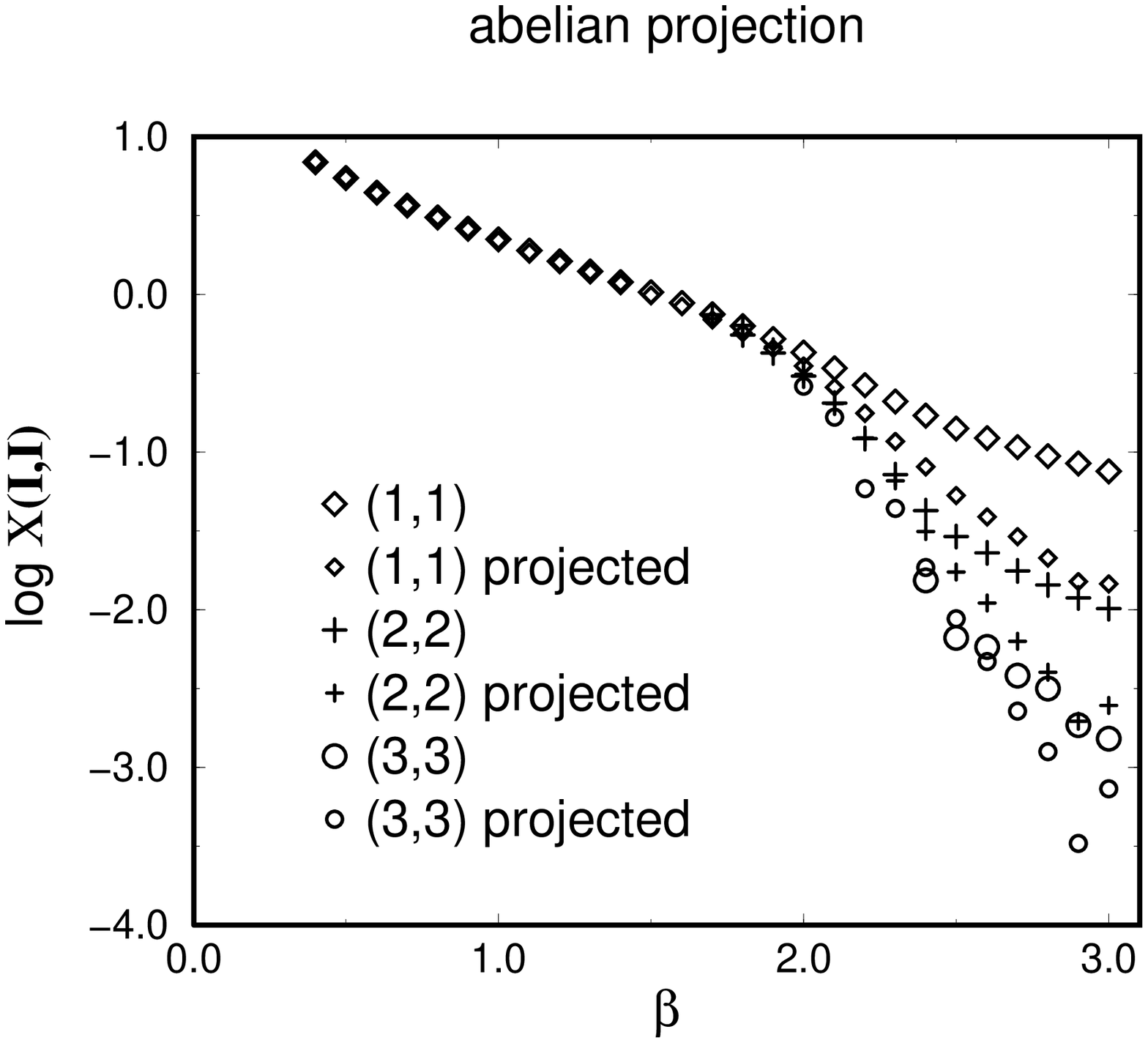} 
\epsfxsize=7cm
\epsffile{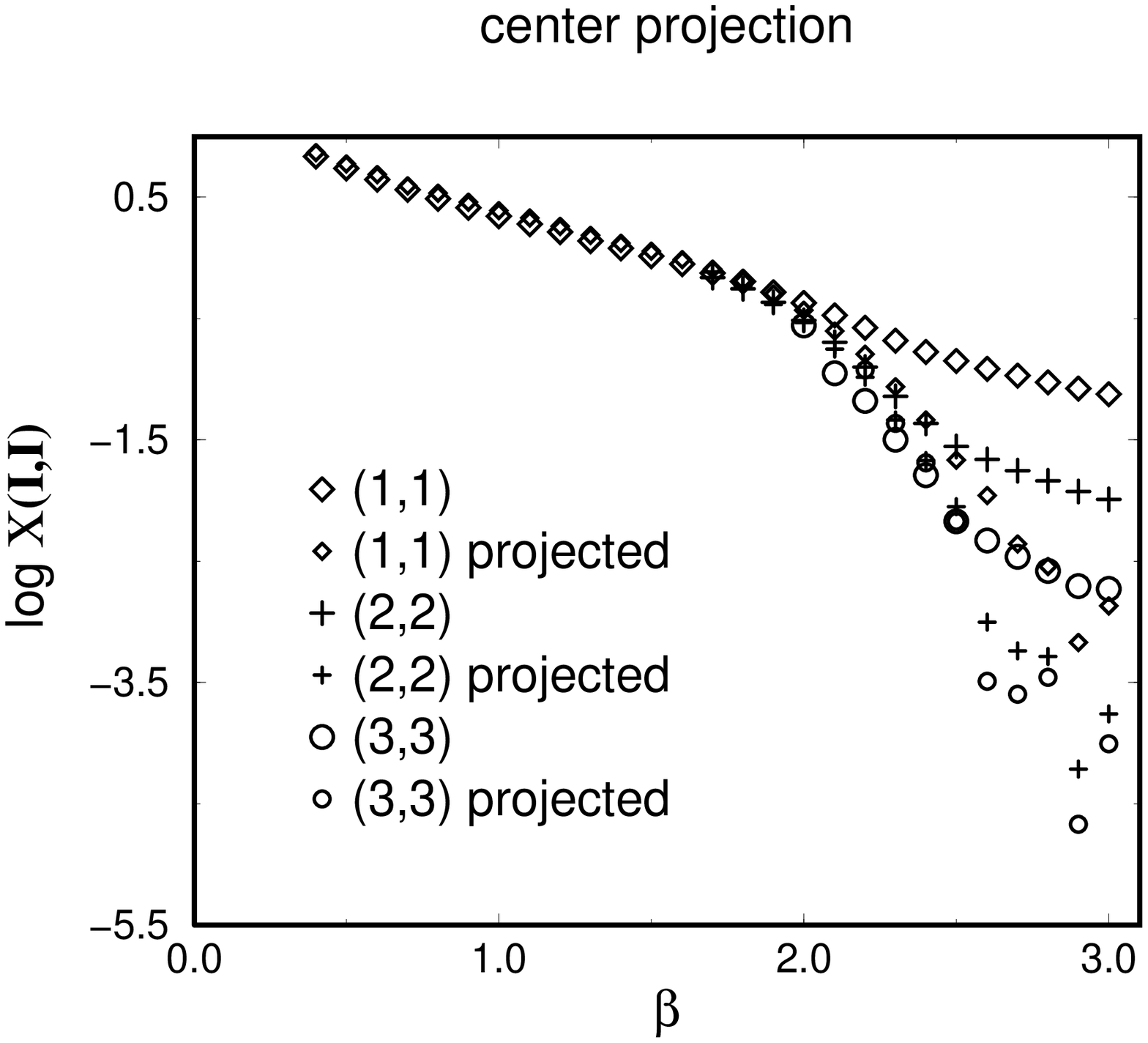} 
}
\vspace{-.8cm} 
\caption{ The Creutz ratios of the full SU(2) gauge theory compared 
   with the values obtained in Abelian projected (left picture) 
   and in center projected (right picture) theory. } 
\label{fig:2} 
\end{figure} 
In figure \ref{fig:2}, we compare the numerical result for the 
Creutz ratios of full SU(2) lattice gauge theory with the result of 
Abelian projected and center projected theory, respectively. 
As observed by many people before~\cite{pol95,deb96}, the string 
tension is almost unaffected by the projection. There is no signal 
of an error of the order of $70\%$ in the string tension as 
one would naively expect from figure 1. 
This shows that the degrees of freedom which are relevant for the 
string tension are largely untouched by the projection mechanism. 

\vskip 0.3cm 
Center projection of lattice gauge theory induces 
magnetic vortices. We follow the definition of Del Debbio et
al.~\cite{deb96}. A plaquette is defined to be part of the 
vortex, if the product of the center projected links which span the 
plaquette is $-1$. As explained in~\cite{deb96}, a (center projected) 
field configuration contributes $(-1)^n$ to the Wilson loop, where 
$n$ is the number of vortices piercing the loop area. In particular, 
it was shown in~\cite{deb96} that the string tension vanishes, 
if the configurations with $n>0$ are discarded. 

\vskip 0.3cm 
Stimulated by these results, we reduce the lattice ground state 
to a medium of vortices with the help of the center projection in order 
to construct a pure vortex vacuum. We then calculate the Creutz 
ratios for several values of $\beta $ employing the vortex model. 
In particular, we measure the probability of finding $n$ vortices which 
pierce the minimal area of the Wilson loop. The Wilson loop $W$ is then 
obtained in the vortex vacuum model by 
\be 
W \; = \; \sum_n (-1)^n \, P(n) \; / \; \sum_n P(n) \; . 
\label{eq:21} 
\en 

\begin{figure}[t]
\parbox{6cm}{ 
\hspace{1cm} 
\centerline{ 
\epsfxsize=6cm
\epsffile{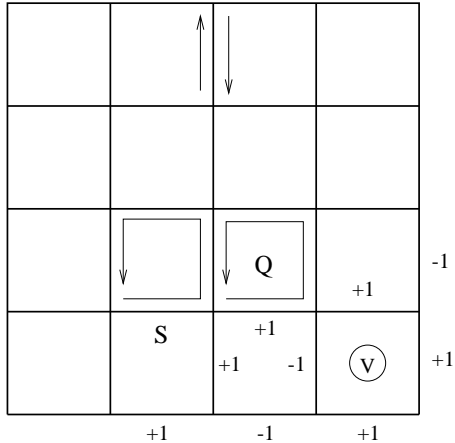} 
}
} \hspace{1cm}
\parbox{7cm}{ 
\caption{ Schematic plot of the contributions of a center projected 
   configuration on the lattice to the Wilson loop. } 
} 
\label{fig:3} 
\end{figure} 
The validity of this formula is easily checked: 
Define $S_x \in \{ \pm 1 \}$ 
to be the center projected link (the index indicating the direction of 
the link is not shown) and $Q_x$ the product of center projected 
links of a particular plaquette at position $x$ (indices suppressed). 
If ${\cal A}$ denotes the minimal area of Wilson loop and $\partial 
{\cal A}$ its boundary, one finds (see figure 3) 
\be 
\prod _{x \in \partial {\cal A} } S_x \; = \; 
\prod _{x \in {\cal A} } Q_x \; = \; (-1)^n \; , 
\label{eq:p}
\en 
where $n$ is the number of vortices piercing the area. 
Here the first equality follows since center projected links $S_x$ 
which do not belong to 
the boundary of the loop appear twice and hence give no contribution 
since $S^2_x=(\pm 1)^2=1$. 

\vskip 0.3cm 
The results for the Creutz ratios calculated from (\ref{eq:21}) 
as function of $\beta $ is identical 
to the results obtained by taking the product of center projected 
links along the boundary of the Wilson loop (small symbols in the 
right picture of figure 2). Hence the vortex vacuum model 
reproduces almost the full string tension as well as the correct 
scaling behavior, if the continuum limit is approached.

\vskip 0.3cm 
\centerline{\bf 3. Vortices scale \hfill } %
\medskip 

\begin{figure}[t]
\centerline{ 
\epsfxsize=8cm
\epsffile{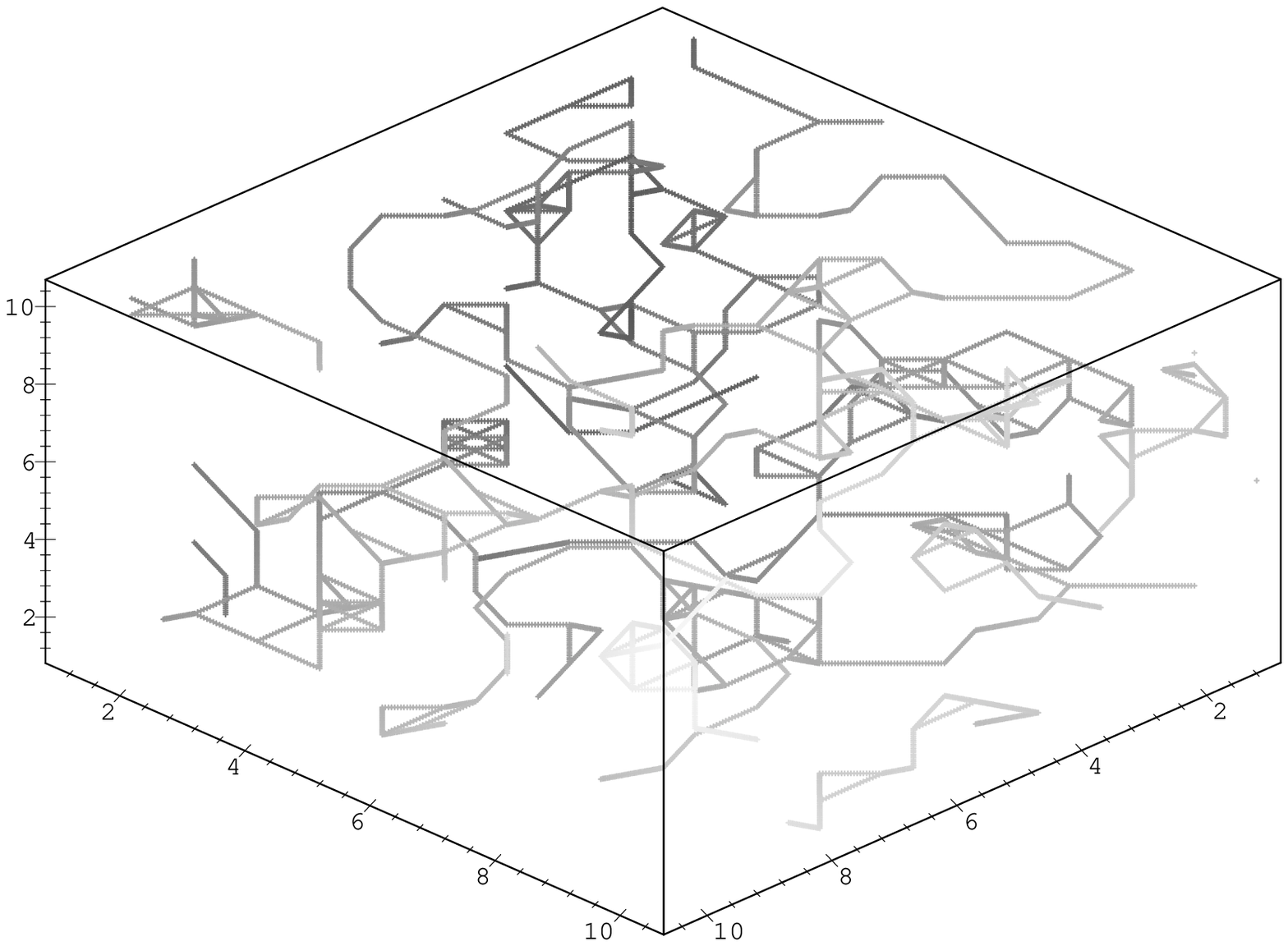} 
\epsfxsize=8cm
\epsffile{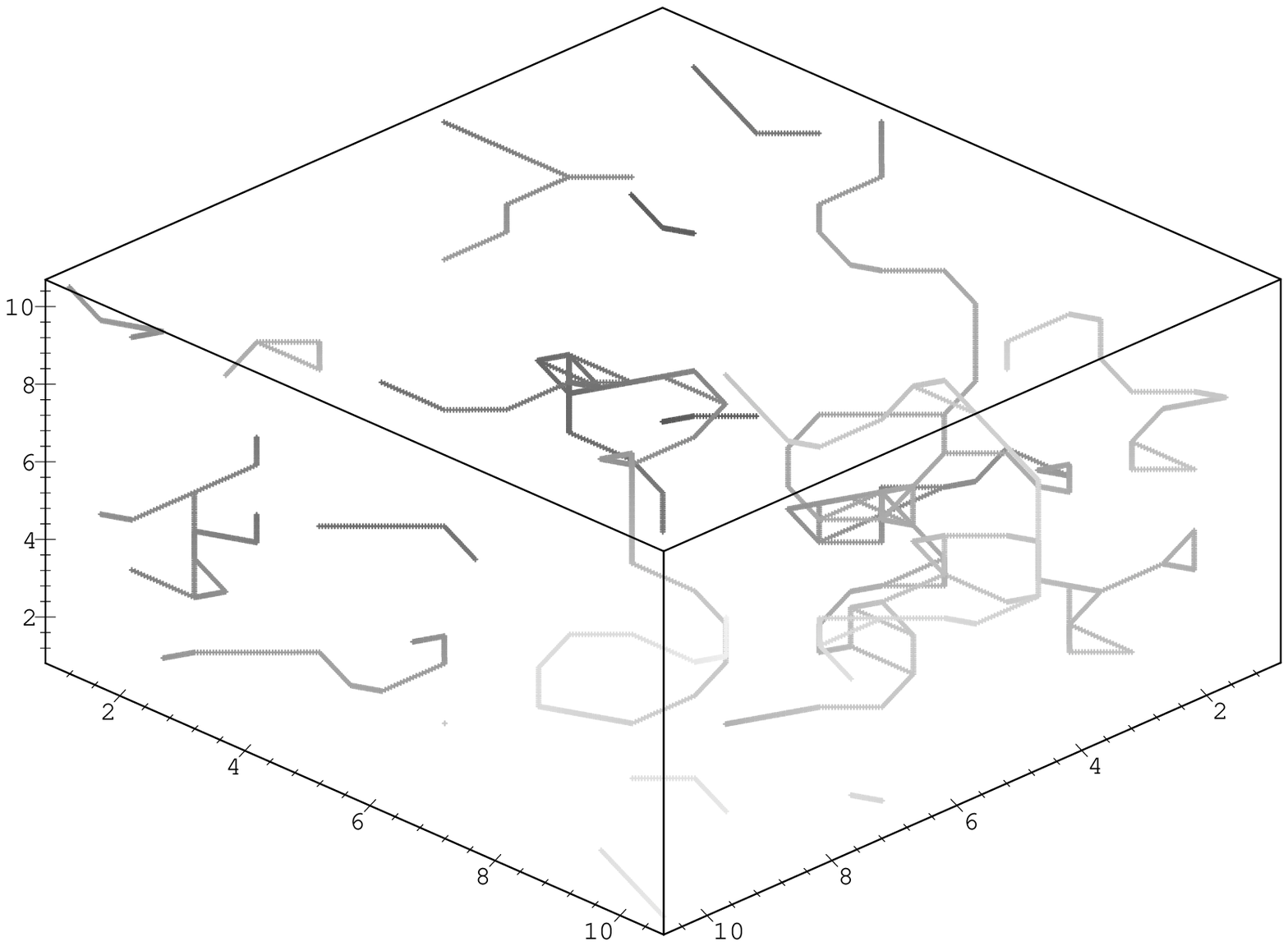} 
}
\vspace{-.8cm} 
\caption{ The distribution of the magnetic vortices in the 
   $10^3$ hypercube of the $10^4$ lattice for a given time slice 
   for $\beta =2.5$ (left) and $\beta =3$ (right). }
\label{fig:4} 
\end{figure} 
In order to get a more explicit picture of the vortex structure 
of the center projected gauge theory, we visualize the vortex 
distribution in coordinate space for a given time slice. In figure 
\ref{fig:4}, we provide two generic configurations of such vortex filled 
state for $\beta =2.5 $ and $\beta =3$. The crucial observation is that 
the gas of vortices is more dilute in the case of $\beta =3$ than 
in the case of $\beta =2.5$. This behavior is anticipated, if we assume 
that the vortex gas is a physical object and therefore renormalization 
group invariant. From the Creutz ratios (see figure \ref{fig:2}), one 
extracts the value $\kappa a^2$, where $\kappa $ is the string tension 
(we use $\kappa \approx (440 \, \hbox{MeV})^2$) and $a$ denotes the 
lattice spacing, 
as function of $\beta $. In practice, we extracted $\kappa a^2$ 
from the Creutz ratios $\chi(3,3)$ using the center projected 
configurations, since there is a clear signal of the perturbative 
renormalization group flow in the center projected theory 
(see figure \ref{fig:2} and also~\cite{deb96}). From $\kappa a^2$ the 
actual value of the lattice spacing $a$ (in units of the string tension) 
is extracted for a given value of $\beta $. 
Larger values of $\beta $ imply a smaller value of the lattice 
spacing due to asymptotic freedom. This implies that we zoom into the 
medium of the vortices, when we go to larger values of $\beta $, if the 
vortices are physical objects (like the string tension). 

\vskip 0.3cm 
In order to establish the physical nature, i.e. the renormalization group 
invariance of the vortex medium, we investigate the average 
number {\cal N} 
of vortices which pierce through an area of $10^2 \, a^2$ as function of 
$\beta $. We then relate the lattice spacing $a$ to the physical 
scale given by the string tension, and extract the density $\rho $ 
of vortices which pierce an area of $1 \, \hbox{fm}^2$. We find that 
$\rho $ is almost independent of $\beta $ in the scaling window 
$\beta \in [2.1,2.4]$. This indicates that $\rho $ is a renormalization group 
invariant and therefore physical quantity. Our results are summarized in 
table below. $L$ denotes the spatial extension of our $10^4$ lattice in 
one direction in physical units. 

\bigskip
\centerline{ 
\begin{tabular}{|c|c|c|c|c|c|} \hline 
$\beta $ & $N$ & $\kappa a^2 $ & $L \; [\hbox{fm}]$ & 
$\rho \; [1/\hbox{fm}^2]$ & $\nu $ \\ 
\hline 
$2.0$ & $22.0 \pm 0.3$ & $0.35$ & $2.7$ & $3.0$ & $3.5$   \\ 
$2.1$ & $18.9 \pm 0.4$ & $0.46$ & $3.1$ & $2.0$ & $3.2$   \\
$2.2$ & $15.1 \pm 0.6$ & $0.40$ & $2.9$ & $1.8$ & $3.0$   \\
$2.3$ & $11.1 \pm 0.4$ & $0.26$ & $2.3$ & $2.0$ & $3.0$   \\
$2.4$ & $8.0  \pm 0.7$ & $0.18$ & $1.9$ & $2.1$ & $2.9$   \\
$2.5$ & $5.8  \pm 0.9$ & $0.11$ & $1.5$ & $2.5$ & $2.6$   \\
\hline 
\end{tabular} 
} 
\bigskip

For $\beta \le 2$, the lattice configurations are far off the continuum 
limit $a \rightarrow 0$, 
whereas the numerical uncertainties in $\kappa a^2$ grow for 
$\beta > 2.5 $. From the numerical results, we estimate 
\be 
\rho \; = \; (1.9 \; \pm \; 0.2) \; \frac{1}{\hbox{fm}^2} \; . 
\label{eq:5} 
\en 

Finally, we extract the number $\nu $ of nearest neighbors which a 
particular point of the vortex has. It is a measure of the branching 
of the vortices. If the vortices form closed loops without branches, 
this number would exactly be two. If the vortices are open strings 
without branches, this number would slightly be smaller than two. 
The table above shows the numerical value of branching value $\nu $ 
for several values of $\beta $. In the scaling window, a value 
$\nu \approx 3$ is consistent with the data, while a significant 
deviation of $\nu $ from $3$ is observed for $\beta \ge 2.5$. This 
deviation is likely due to finite size effects.

\vskip .3cm 
\centerline{ \bf 4. Conclusion  \hfill } 
\medskip 

In this letter, we have studied the vortices arising in center projected 
lattice gauge theory, considered previously in ref.~\cite{deb96}. 
We have evaluated the Creutz ratios in a pure vortex vacuum defined 
via center projection, and have obtained almost the full string tension 
as well as the right scaling behavior towards the continuum limit. 
We have visualized the vortex distribution in coordinate space for 
a given time slice. 
In particular, we have obtained the density $\rho $ of vortices piercing 
the minimal Wilson area as a function of $\beta $. We have observed an 
approximate scaling, which suggests that the vortices are not lattice 
artifacts but physical objects. On the average, we find 
$\rho \approx (1.9 \pm 0.2)/ \hbox{fm}^2 $. 

\vskip .3cm 
Our investigations support the observations of ref.~\cite{deb96} that 
the vortices play the role of ''confiners'' in SU(2) Yang-Mills theory. 

\vskip 1cm 
{\bf Acknowledgment: } 

We thank M.~Engelhardt for helpful discussions and 
useful comments on the \break manuscript.

\end{document}